\documentclass[conference]{IEEEtran}
\IEEEoverridecommandlockouts

\usepackage{cite}
\usepackage{amsmath,amssymb,amsfonts}
\usepackage{algorithm,algpseudocode}
\usepackage{graphicx}
\usepackage{textcomp}
\usepackage{xcolor}
\usepackage{multirow}
\usepackage{booktabs}
\usepackage{float}
\usepackage{url}
\usepackage{bm}
\usepackage{tabularx}
\usepackage{array}
\usepackage{soul}
\usepackage[hidelinks,breaklinks=true]{hyperref}
\usepackage{dsfont}
\usepackage{enumitem}

\newcommand{\Real}{\mathbb{R}}
\newcommand{\Comp}{\mathbb{C}}
\newcommand{\ct}{\dagger}

\begin{document}
\bstctlcite{IEEEexample:BSTcontrol}

\title{Recovering Candidate Circadian Regulators of Arrhythmic Pituitary Hormone Genes Using Reliability-Weighted Magnetic Laplacian with rwMagLap}

\author{
	\IEEEauthorblockN{
		Shabnam Sodagari\thanks{Email: shabnam@csulb.edu}
		and Nick Jasperson
	}
}

\maketitle

\begin{abstract}
We study how to recover candidate circadian-clock regulators of pituitary hormone genes that are important for women's health but do not show a clear 24-hour rhythm in bulk tissue, aiming to nominate clock-linked regulatory targets that could inform future chronopharmacologic and chronotherapeutic strategies. We propose \textbf{rwMagLap}, which builds a graph on rhythmic backbone genes. For each edge, we combine 24-hour fit quality with peak-time phase, represented as a complex unit-circle value, yielding a Hermitian adjacency matrix and a magnetic Laplacian. We insert arrhythmic hormone genes, treated as anchors, by a reliability-weighted nearest-neighbor projection. The projected anchor-neighbor weights are pooled into a soft teleport distribution, and complex personalized PageRank then ranks rhythmic backbone genes by the magnitude of their PageRank scores. In pituitary data, we find that all 11 women's-health anchors are arrhythmic.
Even so, we find that the top-50 list is $7.95\times$ enriched for the 13-gene KEGG circadian set (7 of the 8 set genes in the 454-gene backbone; corrected Benjamini-Hochberg (BH) $p_{\mathrm{BH}}=4\times10^{-6}$) and $4.54\times$ enriched for the 111-gene Reactome set (8 of 16 genes; $p_{\mathrm{BH}}=1.6\times10^{-4}$), while a phase-blind real-valued baseline recovers none. We recover candidates through reliability weighting and phase-aware seeding rather than through magnetic propagation. The magnetic phase adds a different capability: it represents temporal order. On pituitary backbone, the magnetic embedding recovers measured peak-time order of connected pituitary genes with accuracy $0.971$, while $q{=}0$, i.e., no magnetic charge, is at chance.
\end{abstract}


\section{Introduction}
\label{sec:intro}

The body clock circadian timing system regulates daily rhythms, on a 24-hour cycle, in gene expression, hormone secretion, metabolism, and behavior\footnote{This work was supported in part
	by the Doris A. Howell Foundation -- CSUBIOTECH.}. Circadian temporal regulation of the hypothalamic–pituitary–ovarian axis supports gonadotropin release and ovulation. Disruption of circadian organization has been linked to altered ovarian clock function, irregular menstrual cycles, reduced fertility, and adverse pregnancy outcomes, and may interact with endocrine disorders such as polycystic ovary syndrome or PCOS~\cite{Sellix2013,Bedrosian2016}. Women’s reproductive health is therefore closely tied to daily biological timing.

Large cosinor studies~\cite{Ruben2018,Mure2018,Anafi2017} have
measured the rhythmic gene-expression landscape of human tissues. Each
gene gets a peak phase $\phi\in[0,2\pi)$, a relative amplitude rAMP,
and a goodness-of-fit value $R^2$. The phase carries biological meaning:
it tells us when, within the daily cycle, a transcript reaches its
peak. Yet existing graph analyses of circadian
data~\cite{Azadifar2021,Pradhan2012,Matsunaga2009} usually build
\emph{real-valued} co-expression networks, where phase is reduced to a
single similarity score or a yes/no threshold. The standard (real)
Laplacian of such a network is therefore ``phase-blind'': it cannot
represent the cyclic order of peak times.

We also note that many hormone genes of clinical
interest are not rhythmic enough to become nodes in the rhythmic graph.
We call these genes \emph{anchors}: they are biologically motivated
starting points that must be projected onto the rhythmic backbone before
we can rank nearby clock genes. Thus the task is to build a phase-aware
graph on rhythmic genes, insert arrhythmic anchors in a controlled way,
and then rank the rhythmic genes closest to those anchors.

We use a complex-valued graph for the rhythmic backbone genes. For
two rhythmic genes with peak phases $\phi_i$ and $\phi_j$, our proposed method, rwMagLap
stores their phase lag as an antisymmetric angle on the unit circle. The
Hermitian adjacency has phase $\exp\!\left(iq\sin(\phi_i-\phi_j)\right)$, where $q$ is the
magnetic charge. Because $\sin(\phi_j-\phi_i)=-\sin(\phi_i-\phi_j)$, we have
$H_{ji}=\overline{H_{ij}}$. Thus the adjacency matrix is Hermitian
(equal to its own conjugate transpose), which gives a magnetic
Laplacian~\cite{LiebLoss1993,Shubin1994,Fanuel2018,Zhang2021MagNet}.

We weight each link by fit quality, meaning how well each gene's
expression follows a 24-hour cosinor curve, so that poorly fit
oscillations contribute less. We also choose the phase term so that the
graph carries a real ``twist'' (a nonzero accumulated phase) around closed loops, unlike alternative phase choices that cancel around every loop (Section~\ref{sec:gauge}).  We then insert the arrhythmic anchors by a reliability-weighted nearest-neighbor (Nystr\"om) projection (Section~\ref{sec:nystrom}).
The contributions of this paper are thus as follows.
\begin{enumerate}[leftmargin=*,nosep]
\item A reliability-weighted magnetic Laplacian for circadian networks
that uses the phase term $\theta_{ij}=q\sin(\phi_i-\phi_j)$. This phase can accumulate around closed loops, so it cannot be removed by simply rephasing the nodes (Section~\ref{sec:gauge}).
\item A reliability-aware nearest-neighbor projection that places
arrhythmic anchor genes into the network's spectral space, with the
anchor-side $R^2$ applied \emph{after} weight normalization so that it
actually affects the result (Section~\ref{sec:nystrom}).
\item A soft-seeded complex personalized PageRank in which the projected anchor-neighbor weights define the teleport distribution; backbone genes are ranked by the magnitude of their resulting PageRank scores (Section~\ref{sec:complex-ppr}).
\item An enrichment test against two public
circadian gene sets (Section~\ref{sec:enrichment}).
\item A principled evaluation of our magnetic component. Using a
$q{=}0$ control, we separate candidate recovery from temporal ordering. The results show
that our reliability weighting and phase-aware seeding drive the recovery
of candidate genes, while our magnetic charge provides the additional
ability to recover lead--lag order among genes (Sections~\ref{sec:ablation} and~\ref{sec:robustness}).
\end{enumerate}
Thus, rwMagLap produces one pooled ranking over the rhythmic backbone. Each arrhythmic anchor gives reliability-damped weight to nearby rhythmic
genes. These weights form the PageRank seed, and the highest-scoring backbone genes become the final candidates.
Two kinds of data are used. Pituitary transcript data come from
Ruben~\emph{et al.}~\cite{Ruben2018,Ruben2018DataS1}. The validation gene sets
used for enrichment (Section~\ref{sec:enrichment}) come from the
MSigDB~\cite{MSigDBReactomeCirc,MSigDBKeggCirc}.

\section{Related Work}
\label{sec:related}

Genotype-Tissue Expression (GTEx)-based CYCLOPS CYCLOPS (Cyclic Ordering by Periodic Structure) and cosinor analyses identify rhythmic genes across human tissues~\cite{Ruben2018,Anafi2017}. They also provide phase estimates. Primate time-course data show tissue-specific rhythms, including in pituitary~\cite{Mure2018}. Core clock genes are among the broadly rhythmic transcripts. These resources do not model pituitary rhythms as an explicit phase-structured gene graph.

Existing graph approaches use conventional real-valued or unweighted biological graphs. They then apply clique finding, graph clustering, centrality scores, or thresholded correlation filters~\cite{Pradhan2012,Matsunaga2009,Malik2018,Azadifar2021}. These methods do not treat circadian phase as a circular node or edge attribute. They also do not use complex-valued phase weights or phase-aware spectral modes.

Early work on magnetic phases in graph Laplacians studied lattice electrons and discrete magnetic operators on graphs or lattices in a magnetic field~\cite{LiebLoss1993,Shubin1994}. Later applications used magnetic-Laplacian eigenvectors for directed-network visualization~\cite{Fanuel2018} and magnetic-Laplacian graph neural networks for learning on directed graphs~\cite{Zhang2021MagNet}. Related spectral methods have also been developed for signed-network clustering~\cite{Cucuringu2020}. Standard graph signal processing provides the broader spectral framework~\cite{Shuman2013,Ortega2018}, but these surveys do not specifically treat circadian transcriptomic phase as a complex Hermitian graph signal. To our knowledge, magnetic-Laplacian methods have not yet been applied to circadian transcriptomic phase data.

\section{Data and Sources}
\label{sec:data}

\paragraph{Pituitary Cosinor Records}
The database of Ruben~\emph{et al.}~\cite{Ruben2018,Ruben2018DataS1}
contains cosinor regression outputs for 13 GTEx tissues (Aorta, Artery
Coronary, Artery Tibial, Colon, Esophagus, Fat SQ, Fat Visceral, Heart
Atrial, Liver, Lung, Nerve Tibial, Pituitary, Thyroid), with
171{,}413 tissue--gene records in total. The dataset stores
the peak phase in radians on $[0,2\pi)$, so no unit conversion is
needed. We analyze the pituitary to build the graph; for the internal cross-tissue check, we count rhythmicity across all 13 tissue-specific calls, including pituitary (Section~\ref{sec:cross-tissue}). When multiple pituitary records mapped to the same HUGO Gene Nomenclature Committee (HGNC) symbol, we kept
the row with the highest $R^2$, so each gene contributed its best-fitting
cosinor record. After this collapse, the pituitary set has
$N=13{,}227$ transcripts.

The cosinor records provide, for each tissue--gene pair, a false
discovery rate (FDR) for rhythmicity, a peak phase $\phi$, a relative
amplitude (rAMP), and a goodness-of-fit value $R^2$ that summarizes how
well the 24-hour cosinor curve fits the expression profile. Following
Ruben~\emph{et al.}~\cite{Ruben2018}, we call a transcript rhythmic
when $\mathrm{FDR}<0.05$, rAMP$\geq 0.10$, and $R^2\geq 0.10$. In the
pituitary this gives a rhythmic backbone of
$N_{\mathrm{rhy}}=454$ transcripts.

\paragraph{Validation Gene Sets and Matching}
\label{sec:validation-sources}

To check the recovered candidates against known biology, we use two
public circadian gene sets from MSigDB: the 111-symbol
\texttt{REACTOME\_CIRCADIAN\_CLOCK} set~\cite{MSigDBReactomeCirc} and
the 13-symbol \texttt{KEGG\_CIRCADIAN\_RHYTHM\_MAMMAL}
set~\cite{MSigDBKeggCirc}. The validation table can be reconstructed by
downloading these two MSigDB gene sets, keeping the source identifier for each gene, recording whether the identifier is Ensembl or Entrez, and mapping each identifier to the corresponding HGNC symbol using the
identifier fields in the pituitary records. After this mapping, duplicate HGNC symbols within each validation set are
collapsed to one unique symbol before overlap testing, so each validation gene is counted once. For the Reactome set, 120 source Ensembl identifiers collapse to 111 unique HGNC symbols. For the KEGG set, 13 Entrez identifiers map one-to-one to 13 symbols.

The two sets are keyed differently:
\texttt{REACTOME\_CIRCADIAN\_CLOCK}~\cite{MSigDBReactomeCirc} is keyed
on Ensembl gene IDs, while
\texttt{KEGG\_CIRCADIAN\_RHYTHM\_MAMMAL}~\cite{MSigDBKeggCirc} is keyed
on NCBI Entrez gene IDs. The pituitary records carry both Ensembl and
Entrez IDs. A symbol-only match would have missed two of these genes and therefore would have undercounted the overlap.
We therefore match a candidate to a set if its HGNC symbol,
\emph{or} its Ensembl ID, \emph{or} its Entrez ID is in that set. This
matching step is important because it correctly treats ARNTL and BMAL1
as the same gene. With this combined identifier matching, 8 KEGG genes
and 16 Reactome genes fall inside the 454-gene pituitary rhythmic
backbone. 
\paragraph{Anchor Set}
\label{sec:central-observation}
We select our anchor list in Table~\ref{tab:anchor-biology} to consist of the 11 women's-health hormone genes that are present in the pituitary records \cite{Ruben2018DataS1}.
They span the main axes of female reproductive endocrinology~\cite{Melmed2020Williams,YenJaffe2019,Teede2023PCOS}, which is the clinical motivation for the analysis. 
\begin{table}[tbh!]
	\centering
	\small
	\caption{Anchor genes and selected women's-health relevance.}
	\label{tab:anchor-biology}
	\begin{tabularx}{\columnwidth}{@{}l>{\raggedright\arraybackslash}X@{}}
		\toprule
		Gene & Selected women's-health relevance \\
		\midrule
		FSHB   & FSH signaling in folliculogenesis; menstrual-cycle length; menopause timing; fertility and endometriosis associations \\
		LHB    & LH signaling and ovulatory surge; PCOS-related LH dysregulation; hypogonadotropic infertility \\
		GNRHR  & GnRH receptor signaling; pituitary gonadotropin release; ovulation; hypogonadotropic hypogonadism \\
		PRL    & Prolactin signaling; lactation; hyperprolactinemia; ovulatory dysfunction \\
		DRD2   & Dopamine D2 receptor control of prolactin secretion; lactation and hyperprolactinemia physiology \\
		POMC   & ACTH precursor in the stress axis; HPA-axis physiology in pregnancy and postpartum adaptation \\
		CRHR1  & CRH receptor in HPA-axis stress signaling; perinatal stress physiology \\
		TSHB   & TSH/thyroid axis; thyroid function in menstrual regularity, fertility, and pregnancy health \\
		GH1    & Growth hormone; GH--IGF axis; bone density, body composition, and age-related endocrine decline \\
		ESR1   & Estrogen receptor $\alpha$; menopause-related estrogen signaling; endometriosis; hormone-responsive breast cancer \\
		AR     & Androgen receptor; androgen signaling; PCOS-related hyperandrogenism \\
		\bottomrule
	\end{tabularx}
\end{table}

All 11 present anchor genes fail the Ruben rhythmicity filter
(Table~\ref{tab:anchors-arrhythmic}). These are not borderline failures:
the highest $R^2$ is only 0.15 for CRHR1; ESR1 has $R^2=0.019$,
and FSHB and LHB are below 0.005. Thus, the absence of these anchors
from the rhythmic set is unlikely to be explained only by a strict cutoff. A related primate diurnal-transcriptome study shows that rhythmicity varies strongly across bulk tissues, such that genes expressed in multiple tissues may be detected as rhythmic in only a subset of them~\cite{Mure2018}. One likely reason is that bulk pituitary samples mix several endocrine cell types, which can dilute cell-type-specific rhythms. Another possibility is that some endocrine outputs are controlled by pulsatile secretion or physiological stimuli rather than by strong 24-hour rhythms.

\begin{table}[tbh!]
\centering
\caption{Pituitary rhythmicity for the 11 anchor genes related to women's health (highest-$R^2$ row kept). All 11 are arrhythmic.}
\label{tab:anchors-arrhythmic}
\small
\setlength{\tabcolsep}{6pt}
\begin{tabular}{lccc}
\toprule
Anchor & FDR  & $R^2$ & rAMP \\
\midrule
CRHR1 & 0.20 & 0.148 & 0.60 \\
AR    & 0.08 & 0.115 & 0.40 \\
GH1   & 0.54 & 0.081 & 0.57 \\
PRL   & 0.65 & 0.057 & 0.35 \\
TSHB  & 0.59 & 0.047 & 0.46 \\
POMC  & 0.17 & 0.025 & 0.21 \\
ESR1  & 0.82 & 0.019 & 0.12 \\
DRD2  & 0.016& 0.006 & 0.08 \\
GNRHR & 0.87 & 0.005 & 0.19 \\
FSHB  & 0.85 & 0.004 & 0.17 \\
LHB   & 0.48 & 0.001 & 0.05 \\
\bottomrule
\end{tabular}
\end{table}

\section{Method: rwMagLap}
\label{sec:method}

\subsection{Reliability-Weighted Edge Strengths}
Let $i,j \in \{1,\ldots,N_{\mathrm{rhy}}\}$, with
$N_{\mathrm{rhy}}=454$, index the rhythmic backbone genes. Each gene $i$ has
peak phase $\phi_i$, relative amplitude $a_i$, and fit quality $r_i^2$. We define three agreement scores:
\begin{align}
\mathrm{pc}_{ij} &= \cos(\phi_i-\phi_j) \in [-1,1], \label{eq:phase-agreement}\\
\mathrm{ac}_{ij} &= 1 - \tfrac{|a_i - a_j|}{\max(a_i,a_j)} \in [0,1], \label{eq:amplitude-agreement}\\
\mathrm{rw}_{ij} &= \sqrt{r_i^2\cdot r_j^2} \in [0,1]. \label{eq:reliability-weight}
\end{align}
For each possible graph edge $(i,j)$, we define the real-valued link
strength as $w_{ij}=\tfrac{1}{2}(\mathrm{pc}_{ij}+\mathrm{ac}_{ij})\,\mathrm{rw}_{ij}$
when the phase, amplitude, and reliability ($\mathrm{rw}_{ij}$) scores pass the thresholds $\tau_p=\tau_a=0.5$ and $\tau_r=0.10$. Otherwise, we set $w_{ij}=0$, with $w_{ii}=0$. Negative values are also set to zero. 
The Hermitian adjacency has entries
\begin{equation}
	H_{ij}=w_{ij}\exp\!\left(iq\sin(\phi_i-\phi_j)\right),
	\qquad w_{ij}\ge 0 .
\end{equation}
Here $q$ is the magnetic charge. Because
$\sin(\phi_j-\phi_i)=-\sin(\phi_i-\phi_j)$, we have
$H_{ji}=\overline{H_{ij}}$. Thus the adjacency matrix is Hermitian, which gives a magnetic Laplacian~\cite{LiebLoss1993,Shubin1994,Fanuel2018,Zhang2021MagNet}. Since $w_{ij}\geq 0$, this ensures that the normalized magnetic Laplacian in Ea. \eqref{eq:magnetic-laplacian} is positive semidefinite.

\subsection{Removable vs.\ Genuine Magnetic Phase}
\label{sec:gauge}

A natural first choice is to use the phase
$\theta_{ij}^{\mathrm{rem}}=q(\phi_i-\phi_j)$. However, this is not a
useful magnetic choice, because it can be removed by a diagonal unitary
change of basis. Let $W$ be the symmetric real-valued weighted adjacency
matrix of the graph, with entries $w_{ij}$. This choice would define a
complex Hermitian adjacency matrix $H^{\mathrm{rem}}(q)$ with entries
\begin{equation}
	H^{\mathrm{rem}}_{ij}(q)
	=
	w_{ij}e^{iq(\phi_i-\phi_j)} .
	\label{eq:removable-adj-entry}
\end{equation}
Now let
$U(q)=\operatorname{diag}(e^{iq\phi_1},\ldots,e^{iq\phi_{N_{\mathrm{rhy}}}})$.
Then, we can write $
	H^{\mathrm{rem}}(q)=U(q)WU(q)^*,
	\label{eq:removable-unitary}$
because
\begin{align}
	\bigl(U(q)WU(q)^*\bigr)_{ij} &=
	e^{iq\phi_i}w_{ij}e^{-iq\phi_j} =
	w_{ij}e^{iq(\phi_i-\phi_j)} \nonumber\\
	&=
	H^{\mathrm{rem}}_{ij}(q).
	\label{eq:removable-entry-proof}
\end{align}
Thus $H^{\mathrm{rem}}(q)$ is only the real matrix $W$ written in a
rotated complex basis. Its eigenvalues therefore do not depend on $q$.
Because the edge magnitudes $w_{ij}$ are unchanged, the degree matrix is also
unchanged. Since $U(q)$ is diagonal, the corresponding normalized
Laplacian is unitarily equivalent to the real-valued one as well.

The same point can be stated in terms of loop phase. If we follow a cycle in the graph and add the edge phases along the way, a genuine magnetic phase can leave a nonzero total phase around the loop. For the removable choice ($\theta_{ij}^{\mathrm{rem}}=q(\phi_i-\phi_j)$), this does not happen. Each node phase appears once with a plus sign and once with a minus sign, so all terms cancel. Therefore every closed loop has zero accumulated phase, and the construction is only a node rephasing rather than a genuine magnetic twist.

For this reason, in our proposed method, rwMagLap, we use
\begin{equation}
	\theta_{ij}=q\sin(\phi_i-\phi_j), \qquad
	H_{ij}=w_{ij}e^{i\theta_{ij}} .
	\label{eq:hermitian-adj}
\end{equation}
This phase is still antisymmetric, so $H$ remains Hermitian. However, it
is not a simple difference of one node label minus another, so the loop
phase need not cancel on closed cycles. This gives a genuinely magnetic
graph.  
At $q=0$, the operator reduces to the ordinary real-valued
normalized Laplacian. We use $q{=}1.0$ by default. The biological reading
is that $\sin(\phi_i-\phi_j)$ has largest magnitude when two genes peak
about a quarter cycle apart, a lag that can represent possible lead--lag
timing between genes.

\subsection{Normalized Magnetic Laplacian and Spectrum}
We define the degree of node $i$ as
$ d_i=\sum_j |H_{ij}|.$
Since $H_{ij}=w_{ij}e^{i\theta_{ij}}$, this is the same as
$d_i=\sum_j w_{ij}$. With
$D=\mathrm{diag}(d_1,\ldots,d_{N_{\mathrm{rhy}}})$, we use
the normalized magnetic Laplacian~\cite{Fanuel2018,Zhang2021MagNet,vonLuxburg2007}
\begin{equation}
	\mathcal{L}_q = I - D^{-1/2}\,H\,D^{-1/2}.
	\label{eq:magnetic-laplacian}
\end{equation}

Let $\psi_m\in\Comp^{N_{\mathrm{rhy}}}$ denote the $m$-th eigenvector
of $\mathcal{L}_q$, and let $\psi_m(i)$ be its entry for gene $i$.
Let $\Psi=[\,\psi_1\mid\cdots\mid\psi_K\,]$ collect the $K$ retained
eigenvectors, and define the spectral embedding of backbone gene $j$ as
$\mathbf{z}_j=\Psi(j,:)=\big(\psi_1(j),\ldots,\psi_K(j)\big)\in\Comp^K$.
For the lead--lag test in Section~\ref{sec:robustness}, we use the
lowest nontrivial mode $\psi_{\mathrm{dir}}=\psi_2$.

The normalized magnetic Laplacian $\mathcal{L}_q$ is Hermitian because its phase-weighted adjacency matrix $H$ is Hermitian. With nonnegative edge magnitudes, the standard Rayleigh-quotient argument for normalized magnetic Laplacians shows that the eigenvalues of $\mathcal{L}_q$ lie in $[0,2]$. This is the magnetic analogue of the classical normalized-Laplacian spectral bound~\cite{vonLuxburg2007}. For magnetic Laplacians, positive semidefiniteness is established in~\cite{Fanuel2018,Zhang2021MagNet}, and the normalized spectral bound is proved in~\cite{Zhang2021MagNet}. The interval bound is independent of the particular phase choice, although individual eigenvalues may still vary with $q$ and with the edge phases.

\subsection{Inserting Arrhythmic Anchors (Nystr\"om Projection)}
\label{sec:nystrom}

An arrhythmic anchor has no row in $\mathcal{L}_q$. We form a reliability-weighted link distribution from anchor $a$ using only the backbone gene's $R^2$:
\begin{equation}
s_a(j) = \tfrac{1}{2}(\mathrm{pc}_{aj}+\mathrm{ac}_{aj})\sqrt{r_j^2}.
\label{eq:anchor-edge}
\end{equation}
Recall that $R^2$ measures how well that backbone gene fits a 24-hour cosinor curve. We keep the top $K_n=25$ neighbors and rescale their weights to sum to one to obtain $\tilde{s}_a$. In other words, for a backbone gene $j$ over the kept neighbors: $\tilde{s}_a(j)=s_a(j)/\sum_{\ell} s_a(\ell)$. We then apply the anchor's own cosinor $R^2$ to these normalized weights:
\begin{equation}
\tilde{s}_a^{\,(\mathrm{damp})}(j) \;=\; \sqrt{r_a^2}\,\tilde{s}_a(j).
\label{eq:rsq-damping}
\end{equation}
Applying $\sqrt{r_a^2}$ after normalization is important because the
factor would cancel if it were applied before normalization. For
$\tilde{s}_a(j)=s_a(j)/\sum_{\ell} s_a(\ell)$ over the kept neighbors,
pre-normalization damping would have given
\[
\frac{\sqrt{r_a^2}s_a(j)}{\sum_{\ell} \sqrt{r_a^2}s_a(\ell)}
=
\frac{s_a(j)}{\sum_{\ell} s_a(\ell)}
=
\tilde{s}_a(j).
\]
Thus damping before normalization has no effect. By applying the factor
after normalization, we obtain
$\sum_j\tilde{s}_a^{(\mathrm{damp})}(j)=\sqrt{r_a^2}$, so anchors with
smaller cosinor $R^2$ contribute less total seed weight.

The out-of-sample (Nystr\"om \emph{style})
projection~\cite{Nystrom1928,Williams2001Nystrom} then places the
anchor in the spectral space:
\begin{equation}
\hat{\mathbf{c}}_a = \sum_j \tilde{s}_a^{\,(\mathrm{damp})}(j)\,
                              e^{iq\sin(\phi_a-\phi_j)}\,\mathbf{z}_j,
\label{eq:nystrom}
\end{equation}
where $\mathbf{z}_j=\Psi(j,:)=\big(\psi_1(j),\ldots,\psi_K(j)\big)$
is the spectral embedding of backbone gene $j$ defined above.
Because the rows and columns of
$\mathcal{L}_q\in\Comp^{N_{\mathrm{rhy}}\times N_{\mathrm{rhy}}}$
are indexed by genes, each eigenvector has one entry per gene. Hence
$\hat{\mathbf{c}}_a\in\Comp^K$ is the anchor's embedding, formed as the
phase-rotated, weight-averaged combination of its neighbors'
embeddings. The anchor's \emph{coupling strength} is
$\|\hat{\mathbf{c}}_a\|_2$.

The projection therefore yields, for each anchor, both a spectral
embedding $\hat{\mathbf{c}}_a$ whose norm gives its coupling strength
and a reliability-weighted neighbor distribution
$\tilde{s}_a^{(\mathrm{damp})}$ from Eq. ~\eqref{eq:rsq-damping}; we next pool
the latter across all present anchors into a single teleport
distribution and use it to rank the backbone genes by proximity to the anchor gene set.

\subsection{Soft-Seeded Complex Personalized PageRank}
\label{sec:complex-ppr}
Personalized PageRank ranks the nodes of a graph by their proximity to a
chosen starting set under a random walk~\cite{Berkhin2005}. We refer to this starting set, or more generally its associated starting distribution $\mathbf{s}$, as the \emph{seed}: it specifies where the random walk is re-injected when it teleports, and therefore determines the point of view from which the graph is ranked. 
At each step the walk either follows an edge (with probability $\alpha$) or ``teleports'' back to the user-defined starting seed distribution $\mathbf{s}$ (with probability $1-\alpha$). In our simulations, we set edge-following probability to $\alpha = 0.85$, so the teleport probability is $1-\alpha=0.15$.
The stationary distribution $\mathbf{p}$ concentrates mass on
nodes that the seed reaches easily. The teleport vector $\mathbf{s}$ is
where personalization enters: a teleport concentrated on one node ranks
the graph relative to that node, while a teleport spread over a set ranks
relative to the whole set. Standard personalized PageRank uses a
\emph{hard} seed---an indicator that is $1$ on the chosen nodes and $0$
elsewhere. We instead use a \emph{soft} seed: rather than marking the
anchors as in/out, we place a graded, reliability-weighted distribution
over each anchor's backbone neighbors, so a well-fit anchor injects more
teleport mass than a poorly-fit one. In other words, we call our version soft-seeded because the seed is not a single node or a binary set of nodes. Instead, it is a weighted distribution over backbone genes.

We build the teleport distribution by adding the damped weights
from Eq.~\eqref{eq:rsq-damping} over all present anchors to get a starting (teleport) distribution $\mathbf{s}\in\Real_{\geq 0}^{N_{\mathrm{rhy}}}$, then rescale it to sum
to one. Weak anchors contribute less total mass because of the $R^2$-based damping. Each $\tilde{s}_a^{(\mathrm{damp})}$ is supported on anchor $a$'s $K_n$ kept neighbors and zero on the remaining backbone genes, so the sum is a vector over all $N_{\mathrm{rhy}}$ nodes that is nonzero only on the union of the anchors' neighborhoods.
The complex personalized PageRank solution 
\begin{equation}
	\mathbf{p} = (1-\alpha)\,\mathbf{s} + \alpha\,\widetilde{P}^{\,\ct}\,\mathbf{p},
	\qquad \widetilde{P}=D^{-1}H,
\end{equation}
uses the conjugate transpose $\widetilde{P}^{\,\ct}$ of the
row-normalized Hermitian adjacency $\widetilde{P}=D^{-1}H$ (which is
itself not Hermitian, since dividing each row by its degree breaks the
symmetry). Backbone genes are ranked by $|p_i|$. Note that the seed
$\mathbf{s}$ is a set of real, non-negative weights and does not depend
on magnetic charge $q$.
The magnetic charge enters only through the complex propagation operator
$\widetilde{P}^{\,\ct}$, which is why the $q{=}0$ control in
Section~\ref{sec:ablation} isolates the effect of the magnetic phase.

\subsection{Algorithm Summary}
\label{sec:algorithm}

Algorithm~\ref{alg:rwmaglap} gives the full pipeline. The inputs are
the cosinor records and the anchor genes list; the output is a ranking of
every backbone gene. The result is fixed once the charge $q$ and the
teleport probability $1-\alpha$ are fixed.

\begin{algorithm}[t]
\small
\caption{rwMagLap: (\ul{r}eliability-\ul{w}eighted \ul{Mag}netic \ul{Lap}lacian) for recovering circadian-clock neighbors of arrhythmic anchor genes.}
\label{alg:rwmaglap}
\begin{algorithmic}[1]
\Require Cosinor records $\{(\phi_i, a_i, r_i^2)\}_{i=1}^{N}$;
         anchor list $\mathcal{A}$; thresholds $\tau_p,\tau_a,\tau_r$;
         charge $q$; teleport $1{-}\alpha$; eigen-rank $K$;
         neighbors $K_n$
\State Keep rhythmic backbone $\mathcal{B}$ via FDR, rAMP, $R^2$
\State $w_{ij}\gets \tfrac{1}{2}(\mathrm{pc}_{ij}{+}\mathrm{ac}_{ij})\,\mathrm{rw}_{ij}\,
\mathds{1}\!\left[\mathrm{pc}_{ij}{\geq}\tau_p,\,\mathrm{ac}_{ij}{\geq}\tau_a,\,\mathrm{rw}_{ij}{\geq}\tau_r\right]$
\Comment{scores in \eqref{eq:phase-agreement}--\eqref{eq:reliability-weight}}
\State $H_{ij}\!\gets\! w_{ij}\,e^{iq\sin(\phi_i-\phi_j)}$
       \Comment{Hermitian adjacency, \eqref{eq:hermitian-adj}}
\State $D\!\gets\!\mathrm{diag}\bigl(\textstyle\sum_j |H_{ij}|\bigr)$;\ \ 
       $\mathcal{L}_q\!\gets\! I - D^{-1/2} H D^{-1/2}$
\State $(\lambda_{1:K},\Psi)\!\gets\!\mathrm{eigh}(\mathcal{L}_q,\,K)$
\State $\mathbf{s}\!\gets\!\mathbf{0}\in\Real_{\geq 0}^{|\mathcal{B}|}$
       \Comment{seed distribution}
\ForAll{anchor $a\in\mathcal{A}$}
  \State $s_a(j)\!\gets\!\tfrac{1}{2}(\mathrm{pc}_{aj}{+}\mathrm{ac}_{aj})\sqrt{r_j^2}$
  on top-$K_n$ neighbors
  \State $\tilde{s}_a\!\gets\!s_a/\|s_a\|_1$
  \State $\tilde{s}_a^{(\mathrm{damp})}\!\gets\!\sqrt{r_a^2}\,\tilde{s}_a$
  \Comment{post-norm.\ damping, \eqref{eq:rsq-damping}}
  \State $\mathbf{s}\!\gets\!\mathbf{s} + \tilde{s}_a^{(\mathrm{damp})}$
\EndFor
\State $\mathbf{s}\!\gets\!\mathbf{s}/\|\mathbf{s}\|_1$
\State Solve $\mathbf{p}\!=\!(1{-}\alpha)\mathbf{s}\!+\!\alpha\,\widetilde{P}^{\ct}\mathbf{p}$, 
       $\widetilde{P}\!=\!D^{-1}H$, by power iteration
\State \Return $|\mathbf{p}|$ sorted high to low
\end{algorithmic}
\end{algorithm}

The same neighbor count $K_n=25$ and eigen-rank $K=24$ are used in every
step, including the null tests (random-backbone and phase-shuffle controls) in Section~\ref{sec:cross-tissue}.

\subsection{Computational Complexity}
\label{sec:complexity}

For a backbone of $N_{\mathrm{rhy}}$ genes, the main costs are: building the adjacency in
$\mathcal{O}(N_\mathrm{rhy}^2)$; finding the $K$ smallest eigenpairs in
$\mathcal{O}(K N_{\mathrm{rhy}}^2)$ with iterative Hermitian
solvers~\cite{vonLuxburg2007}; and computing Personalized PageRank by power
iteration in $\mathcal{O}(T(|V|+|E|))$, or $\mathcal{O}(T|E|)$ when edge scans
dominate, for $T$ iterations~\cite{Berkhin2005}. For pituitary backbone, we have $N_\mathrm{rhy}=454$, number of edges
$|E|=46{,}005$, $K=24$, $T\!\approx\!24$.

\section{Results}
\label{sec:results}

\subsection{Spectrum and Anchor Coupling}

Our phase-aware spectral clustering on the magnetic embedding partitions the 454 backbone genes into ten phase-coherent modules (genes grouped by similar peak time), of sizes $\{17,33,40,40,44,49,55,58,58,60\}$, which sum to 454. The clustering acts on the \emph{spectral embedding}---each gene's coordinates in the lowest nontrivial eigenvectors of $\mathcal{L}_q$ (those with the smallest eigenvalues), where genes with similar peak time
lie close together. 
The module count of ten maximizes the mean \emph{silhouette} of this embedding, a standard cluster-quality score that is higher when groups are more cleanly separated. 
The module count was selected by maximizing the mean \emph{silhouette}
score over candidate counts $\{2,\ldots,15\}$ using the same
phase-clustering embedding. The selected partition contains ten modules.
The partition only summarizes phase organization; it does not enter the candidate ranking, which is computed on $H$ and is independent of the number of modules.

Table~\ref{tab:anchor-coupling} lists the 11 anchors by
coupling strength. The values follow the after-normalization
$\sqrt{r_a^2}$ damping: the weakest-fit anchors also have the smallest couplings: LHB (luteinizing-hormone surge anchor), FSHB (follicle-stimulating hormone), GNRHR (GnRH receptor), and DRD2 (dopaminergic suppression of prolactin).
The largest coupling goes to CRHR1 (CRH receptor; HPA stress axis), which also has the highest $R^2$. 
CRHR1, AR (androgen receptor; PCOS/hyperandrogenism), and TSHB
(TSH; thyroid--fertility axis) all couple to partners from the same
peak-time backbone neighborhood, but their exact top-3 sets differ:
CRHR1 couples most strongly to PER3, DBP, and C1orf51; AR to PER3, TEF,
and NR1D2; and TSHB to TEF, NR1D2, and PER3. GH1 (growth hormone; bone
density) instead couples to the antiphase positive-arm cluster
(ARNTL/BMAL1, NPAS2, CDKN1A), while DRD2 shows lower coupling and a
distinct neighbor set (MGST2, MKKS, COMMD9).

\begin{table}[tbh!]
	\centering
	\caption{Magnetic-Laplacian coupling of the 11 anchor genes and
		their top-3 phase-locked backbone partners.
	}
	\label{tab:anchor-coupling}
	\setlength{\tabcolsep}{3pt}
	\footnotesize
	\begin{tabular}{l@{\hspace{6pt}}c@{\hspace{6pt}}c@{\hspace{6pt}}l}
		\toprule
		Anchor & $R^2$ & $\|\hat{\mathbf{c}}_a\|_2$ & Top-3 phase-locked partners \\
		\midrule
		CRHR1 & 0.148 & 0.066 & PER3, DBP, C1orf51 \\
		AR    & 0.115 & 0.054 & PER3, TEF, NR1D2 \\
		GH1   & 0.081 & 0.046 & ARNTL, NPAS2, CDKN1A \\
		PRL   & 0.057 & 0.036 & SLC25A21-AS1, KBTBD7, TTC30B \\
		TSHB  & 0.047 & 0.034 & TEF, NR1D2, PER3 \\
		POMC  & 0.025 & 0.018 & CRY2, PER3, C1orf51 \\
		ESR1  & 0.019 & 0.011 & TEF, CCDC115, MICU1 \\
		DRD2  & 0.006 & 0.010 & MGST2, MKKS, COMMD9 \\
		GNRHR & 0.005 & 0.007 & ZNF174, RAD17, BPNT1 \\
		FSHB  & 0.004 & 0.005 & TEF, CCDC115, SLC37A4 \\
		LHB   & 0.001 & 0.004 & PER1, CRY2, EMD \\
		\bottomrule
	\end{tabular}
\end{table}

\subsection{Recovering Top-Ranked Clock-Gene Neighbors of Anchors: rwMagLap vs. Phase-Blind Baseline}
rwMagLap scores each backbone gene with a random walk that restarts from the anchors' nearest rhythmic neighbors and ranks genes by how often the walk reaches them (Section~\ref{sec:complex-ppr}). 
Table~\ref{tab:top-candidates} shows the top-15 from our proposed rwMagLap method and phase-blind baseline. Seven of rwMagLap’s top-15 are canonical or clock-associated circadian genes, including activators and repressors from the core transcriptional feedback network: PER3, DBP, BHLHE41, ARNTL/BMAL1, CRY2, PER2, and NPAS2. The phase-blind baseline's top-15 contains none. 

Because the personalized-PageRank seed pools all 11 anchors
(Section~\ref{sec:complex-ppr}), each candidate backbone circadian gene
reflects proximity to the anchor set as a whole, not to a single anchor
(per-anchor neighbors appear in Table~\ref{tab:anchor-coupling}).
Consistent with the phase clusters above, anchors
CRHR1, AR, and TSHB seed most of the list, whereas GH1 seeds the antiphase positive-arm candidates ARNTL/BMAL1 and NPAS2---so both clock arms are recovered because different anchors seed different arms.

\begin{table}[tbh!]
\centering
\caption{Top-15 candidates from each method. $\bullet$ = present in
\texttt{REACTOME\_CIRCADIAN\_CLOCK} or
\texttt{KEGG\_CIRCADIAN\_RHYTHM\_MAMMAL} (matched by symbol, Ensembl,
or Entrez ID)~\cite{MSigDBReactomeCirc,MSigDBKeggCirc}.}
\label{tab:top-candidates}
\footnotesize
\setlength{\tabcolsep}{4pt}
\begin{tabular}{c|l|l}
\toprule
Rank & Proposed (rwMagLap) & Phase-blind baseline \\
\midrule
1  & TEF             & C17orf75    \\
2  & TMEM80          & NAPEPLD     \\
3  & \textbf{PER3}$\bullet$  & FAM117B     \\
4  & NR1D2           & SLC25A35    \\
5  & \textbf{DBP}$\bullet$  & TARBP1      \\
6  & \textbf{BHLHE41}$\bullet$ & LANCL2   \\
7  & C1orf51         & FAM47E      \\
8  & MDM1            & CARF        \\
9  & \textbf{ARNTL}$\bullet$ (BMAL1) & TOLLIP-AS1 \\
10 & CCDC136         & FMN2        \\
11 & \textbf{CRY2}$\bullet$  & METTL25     \\
12 & \textbf{PER2}$\bullet$  & NAV3        \\
13 & COL21A1         & DNAJC9-AS1  \\
14 & HLF             & DUS4L       \\
15 & \textbf{NPAS2}$\bullet$ & CEP72       \\
\bottomrule
\end{tabular}
\end{table}
The phase-blind baseline fails because no anchor is rhythmic at the bulk pituitary level (Table~\ref{tab:anchors-arrhythmic}), so it has nothing to seed from inside the backbone and is seeded uniformly, returning a nearly flat ranking. rwMagLap avoids this by placing arrhythmic anchors into the spectral space through the phase-aware projection. This is a substantive distinction about seeding, and it is separate from the magnetic-charge parameter.
\subsection{Enrichment Against Reference Circadian Gene Sets}
\label{sec:enrichment}

We test whether the top-50 candidates overlap two public circadian datasets~\cite{MSigDBKeggCirc,MSigDBReactomeCirc} more than expected by chance using a hypergeometric test. The universe is the 454-gene backbone, and we apply Benjamini--Hochberg (BH) correction across the four method--set tests. Table~\ref{tab:enrichment} reports the resulting fold enrichments.

\begin{table}[tbh!]
\centering
\caption{Overlap of the top-50 candidates with the two public circadian
sets~\cite{MSigDBReactomeCirc,MSigDBKeggCirc}. $k$: observed overlap; $K_{\mathrm{eff}}$: set genes inside the
454-gene backbone; $E_{\mathrm{exp}}$: expected overlap by chance. Matching is by
symbol, Ensembl, or Entrez ID. The phase-blind baseline recovers zero
from either set.}
\label{tab:enrichment}
\small
\setlength{\tabcolsep}{5pt}
\begin{tabular}{llccccc}
\toprule
Method & Set & $k$ & $K_{\mathrm{eff}}$ & $E_{\mathrm{exp}}$ & Fold & BH $p_{\mathrm{BH}}$ \\
\midrule
rwMagLap   & Reactome & 8 & 16 & 1.76 & 4.54$\times$ & $1.6{\times}10^{-4}$ \\
rwMagLap   & KEGG    & 7 &  8 & 0.88 & \textbf{7.95$\times$} & $\bm{4{\times}10^{-6}}$ \\
Phase-blind& Reactome & 0 & 16 & 1.76 & 0.00 & 1.00 \\
Phase-blind& KEGG   & 0 &  8 & 0.88 & 0.00 & 1.00 \\
\bottomrule
\end{tabular}
\end{table}
As in Table~\ref{tab:enrichment}, rwMagLap top-50 recovers 7 of the 8 KEGG genes in the backbone and 8 of the 16 Reactome genes from a 50-out-of-454 ranking; chance would give only $0.88$ and $1.76$, respectively. This corresponds to $7.95\times$ KEGG enrichment ($p_{\mathrm{BH}}=4\times10^{-6}$) and $4.54\times$ Reactome enrichment ($p_{\mathrm{BH}}=1.6\times10^{-4}$). The phase-blind baseline recovers zero from either set: with no anchor in the backbone, it is seeded uniformly and returns a nearly flat ranking.

Our two design choices produce this:
reliability-weighted link strengths, which prevent poorly-fit genes from
adding noise; and a phase-aware seed that places each arrhythmic anchor next to backbone genes with a similar peak time and similar amplitude.

\subsection{Where the Recovery Comes From: the $q{=}0$ Control}
\label{sec:ablation}

The natural question is whether the magnetic phase is responsible for
the recovery. To test this, we set the charge to $q=0$. This keeps the reliability-weighted links, the degrees, and the real seed exactly
the same; the only thing that changes for $q>0$ is the complex phase on
the links during propagation. So $q=0$ versus $q>0$ isolates the
contribution of the magnetic phase.

\begin{figure}[tbh!]
\centering
\includegraphics[width=\columnwidth]{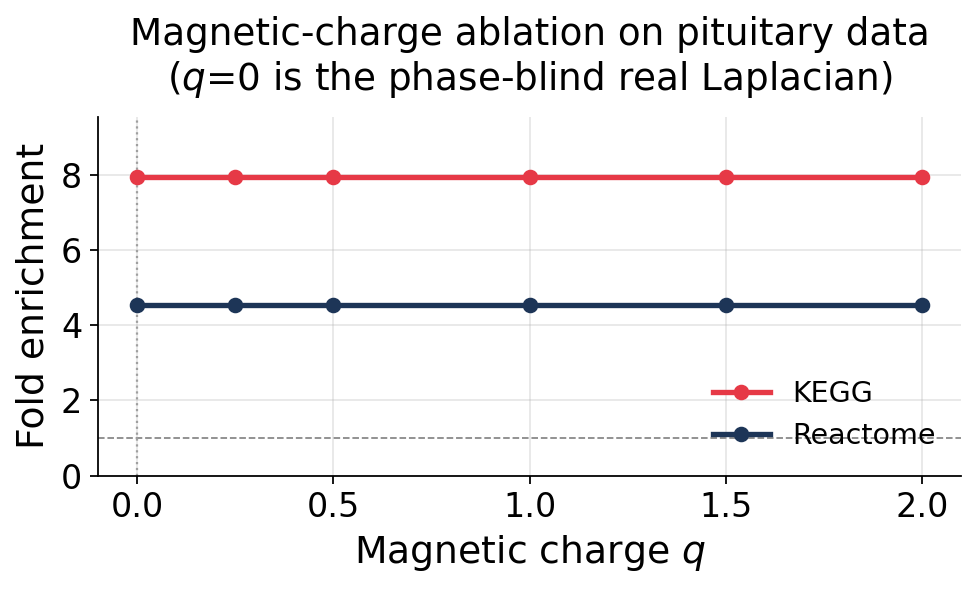}
\caption{Enrichment of the top-50 as the magnetic charge $q$ varies.
$q=0$ is the plain real Laplacian (same pipeline otherwise). Both
curves are flat, so the role of the magnetic charge is not which genes are
recovered, but the temporal order of the genes (Section~\ref{sec:robustness}).}
\label{fig:ablation}
\end{figure}

The answer is in Fig.~\ref{fig:ablation}: the enrichment is \emph{identical at every
charge from $0$ to $2$}. The same 7 KEGG genes and the same 8 Reactome
genes are recovered, with fold values $7.95\times$ and $4.54\times$
throughout. The top-25 lists overlap by Jaccard similarity $\geq 0.92$ between
$q=0$ and $q=1$. In other words, a plain real-valued Laplacian with the
same reliability weighting and the same phase-aware seed recovers the
clock genes just as well. The recovery is produced by the
reliability weighting and the phase-aware seeding, not by the magnetic
charge. Next, we show that the magnetic charge is needed to encode temporal ordering of genes.

\subsection{Effect of the Magnetic Charge on the Ranking and Directionality on the Pituitary Backbone}
\label{sec:robustness}

We swept $q\in\{0,0.25,0.5,1.0,1.5,2.0\}$ and compared each
top-$25$ candidate set with the top-$25$ set obtained at the default
value $q=1.0$. The minimum Jaccard similarity over the tested grid
was $0.786$, while the lists for $q=0, 0.25, 0.5$ already
overlapped strongly with the default $q=1.0$ list. This agrees with
the enrichment control above: changing $q$ has little effect on the
magnitudes $|p_i|$ that set the candidate ranking.

The enrichment test only asks \emph{which} genes are recovered; it is
blind to their temporal \emph{order}. 
To see what the magnetic charge contributes, we run a check on the pituitary backbone. Here the
reference order is the measured cosinor peak phase of each rhythmic
gene. 
For each connected ordered gene pair $(i,j)$, let
$\eta_i=\psi_{\mathrm{dir}}(i)$ and
$\eta_j=\psi_{\mathrm{dir}}(j)$ be the entries of the lowest nontrivial
eigenvector. We use
$\operatorname{Im}\!\left(\eta_i\overline{\eta_j}\right)
=
|\eta_i||\eta_j|\sin\!\left(\arg\eta_i-\arg\eta_j\right)$
to predict local lead--lag order. We compare
$\operatorname{sign}\!\left(\operatorname{Im}(\eta_i\overline{\eta_j})\right)$
with $\operatorname{sign}\!\left(\sin(\phi_i-\phi_j)\right)$.
Since retained
edges satisfy $\mathrm{pc}_{ij}=\cos(\phi_i-\phi_j)\ge 0.5$, connected genes differ
in peak phase by at most $60^\circ$. Thus the test compares only local
lead--lag relations between nearby phases, avoiding ambiguity from the
circular $24$-hour phase wraparound.

\begin{figure}[tbh!]
\centering
\includegraphics[width=\columnwidth]{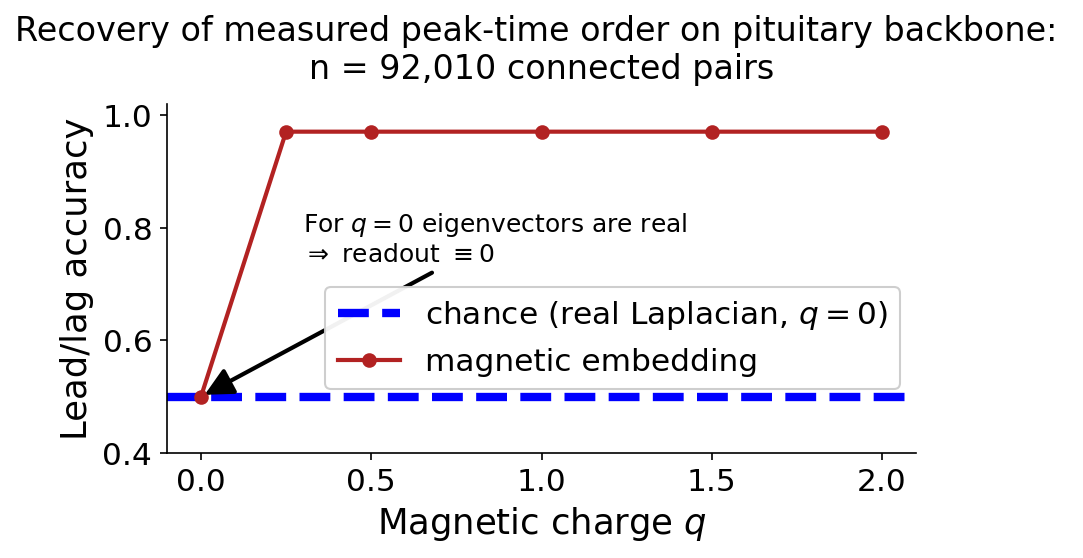}
\caption{Lead/lag direction accuracy in pituitary backbone for connected backbone pairs, using measured cosinor peak phases as the reference order. At $q=0$ the real Laplacian's eigenvectors are real, so
$\mathrm{Im}(\eta_i\overline{\eta_j})$ is identically zero and accuracy is exactly chance ($0.5$). For $q>0$, the lowest magnetic eigenvector
recovers the measured peak-time order at $0.971$ (bootstrap $95\%$ CI over the $92{,}010$ connected ordered pairs).}
\label{fig:real-direction}
\end{figure}

Across the $46{,}005$ backbone edges ($92{,}010$ ordered pairs; mean
phase gap $23.5^\circ$), the accuracy is exactly $0.500$ at $q=0$ and
$0.971$ (bootstrap $95\%$ confidence interval (CI) $[0.969,0.972]$) for every $q\ge0.25$
(Fig.~\ref{fig:real-direction}). The $q=0$ value is expected: the real
Laplacian has real eigenvectors, so $\mathrm{Im}(\eta_i\overline{\eta_j})$
is identically zero. Thus the magnetic embedding captures the measured
peak-time order of connected pituitary genes, while the real-valued
operator cannot encode that order. Fig.~\ref{fig:real-direction}
shows that the magnetic embedding preserves the directional phase structure already present in the real graph, whereas the $q=0$ operator
cannot represent that structure. Thus, the reliability-weighted, phase-seeded graph
selects \emph{which} genes are close to the anchors
(Section~\ref{sec:ablation}), while the magnetic charge encodes \emph{their order}. 

\subsection{Internal Cross-Tissue Check}
\label{sec:cross-tissue}

As an internal validation using only the original dataset~\cite{Ruben2018DataS1}, we asked whether the top-50 candidates are rhythmic in more tissues than expected for arbitrary genes from the pituitary rhythmic backbone. For each candidate gene, we counted the number of the 13 tissues in which that gene passed the rhythmicity filter. We then averaged this count over the 50 candidates to obtain $\bar n_{\mathrm{tis}}$. To form a size-matched random-backbone null, we sampled 50 genes from the 454-gene pituitary backbone without replacement and computed the same mean tissue count. We repeated this procedure 500 times. The rwMagLap top-50 is rhythmic in $5.46$ tissues on average, versus $3.75$ for the random backbone samples; none of the 500 random samples reached this value (Table~\ref{tab:cross-tissue}).

\begin{table}[tbh!]
	\centering
	\caption{Cross-tissue check using only the~\cite{Ruben2018DataS1} dataset.
		$\bar n_{\mathrm{tis}}$: mean number of tissues (of 13) in which a
		top-50 candidate is rhythmic.}
	\label{tab:cross-tissue}
	\small
	\setlength{\tabcolsep}{4pt}
	\begin{tabular}{lcc}
		\toprule
		Method & $\bar n_{\mathrm{tis}}$ & $\geq 3$ tissues \\
		\midrule
		rwMagLap top-50           & \textbf{5.46} & \textbf{78\%} \\
		Phase-blind top-50        & 3.50  & 64\% \\
		Random backbone ($\times 500$) & 3.75 & --- \\
		\bottomrule
	\end{tabular}
\end{table}

We also used a phase-shuffle null to ask whether this cross-tissue signal depends on the exact pituitary phase order. After shuffling pituitary phases and rerunning the ranking, the null mean was $5.34$, close to the observed value $5.46$. Thus, the cross-tissue signal is not driven by the exact pituitary phase order, consistent with the $q=0$ control. The magnetic charge instead contributes temporal ordering--which gene peaks before which--rather than candidate selection.

\subsection{Clinical Value}
\label{sec:chronotherapy}
The clinical value is not only that rwMagLap recovers known clock genes,
but also that it suggests a phase-coherent neighborhood
for each \emph{arrhythmic} hormone anchor
(Table~\ref{tab:anchor-coupling}). For each anchor, the top-3 partners
are candidate \emph{phase markers}---transcriptomic neighbors that share
a peak time---for that anchor's pharmacological target. For example,
CRHR1 couples most strongly to PER3 and DBP. 
These partners are testable hypotheses for future timing experiments, not inferred dosing windows.
The method remains usable for the large class of
pharmacologically important genes whose own transcripts are too weak or too on-demand to register as rhythmic in bulk tissue---common in women's
reproductive endocrinology. A natural follow-up is a dose--timing experiment on an anchor target (e.g.\ AR or CRHR1) against its phase-locked partners.

\section{Conclusion}
\label{sec:conclusion}

We presented \textbf{rwMagLap}, a reliability-weighted
magnetic-Laplacian method for phase-aware graph analysis of circadian
transcriptomes, applied to arrhythmic women's-health pituitary hormone
genes (called anchors). From anchors that are all arrhythmic in bulk tissue, the method
recovers a top-50 list that is $7.95$ times enriched in the KEGG
circadian set ($p_{\mathrm{BH}}=4\times10^{-6}$; 7 of 8 backbone set genes) and
$4.54$ times in the Reactome set ($p_{\mathrm{BH}}=1.6\times10^{-4}$), while a phase-blind baseline recovers none. Through testing a zero magnetic charge control we show that the recovery comes from the reliability
weighting and the phase-aware seeding. What the magnetic charge uniquely adds is recovery of the temporal order of
genes, which a real-valued graph cannot represent.
The practical message is twofold: a reliability-weighted, phase-seeded graph
is an effective way to find clock neighbors of arrhythmic hormone genes;
and a magnetic Laplacian is best used for what it alone can do---reading
the direction of circadian phase.
By recovering canonical clock-gene neighbors for arrhythmic anchors, rwMagLap identifies candidate \emph{phase markers}---transcriptomic neighbors with aligned peak times---that can guide follow-up chronobiology and dose--timing experiments, offering an early discovery step toward experimentally testable chronotherapy hypotheses.
Future work should test single-cell pituitary data and sex-stratified cohorts, given reported sex- and age-dependent organization of human circadian transcriptomes~\cite{Talamanca2023}.

\bibliographystyle{IEEEtran}
\bibliography{refs}

\end{document}